\documentclass[a4paper,12pt,preprint,amsmath,showpacs,
nofootinbib,superscriptaddress]{revtex4}
\usepackage{mathrsfs}
\usepackage{amssymb}
\usepackage{amsmath}
\usepackage{graphicx}
\usepackage{multirow}
\usepackage{subfigure}
\usepackage{float}
\usepackage{graphics}

\def\as{\alpha_s}

\def\nno{\nonumber\\}
\newcommand{\be}{\begin{eqnarray}}
\newcommand{\ee}{\end{eqnarray}}

\def\ses{\sqrt{S}=7~\text{TeV}}
\def\eis{\sqrt{S}=8~\text{TeV}}
\def\ths{\sqrt{S}=13~\text{TeV}}
\def\fos{\sqrt{S}=14~\text{TeV}}
\def\wz{W^{\pm}Z}
\def\mv{M_{VZ}}

\begin{document}
\title{Threshold Resummation for $\wz$ and $ZZ$ Pair Production at the LHC}
\author{Yan  Wang}
\affiliation{Department of Physics and State Key Laboratory of
Nuclear Physics and Technology, Peking University, Beijing 100871,
China}
\author{Chong Sheng Li}
\email{csli@pku.edu.cn}
\affiliation{Department of Physics and State Key Laboratory of
Nuclear Physics and Technology, Peking University, Beijing 100871,
China}
\affiliation{Center for High Energy Physics, Peking University, Beijing, 100871, China}
\author{Ze Long Liu}
\affiliation{Department of Physics and State Key Laboratory of
Nuclear Physics and Technology, Peking University, Beijing 100871,
China}
\author{Ding Yu Shao}
\affiliation{Department of Physics and State Key Laboratory of
Nuclear Physics and Technology, Peking University, Beijing 100871,
China}



\begin{abstract}
We perform the threshold resummation for $W^{\pm}Z$ and $ZZ$ pair production at the  next-to-next-to-leading logarithmic accuracy in Soft-Collinear Effective Theory at the LHC. Our results show that the resummation effects increase the total cross sections by about 7\% for $ZZ$ production and 12\% for $\wz$ production with $\sqrt{S}= 7,~8,~13$ and $14$ TeV, respectively, and the scale uncertainties are significantly reduced. Besides, our numerical results are well consistent with experimental data reported by the ATLAS and CMS collaborations.
\end{abstract}

\maketitle

\section{Introduction}\label{s1}
The gauge boson pair production plays an important role at the LHC, not only for testing the $SU(2)\times U(1)$ gauge structure of the Standard Model (SM), but also for understanding the background of the SM Higgs and the new physics signal.
In particular, $ZZ$ production is the irreducible background process to the SM Higgs boson signal. Besides, the $WWZ$ tri-linear coupling, which can be used to search for heavy charged gauge boson, is constrained by the $\wz$ production.
Deviations between the measured data and the SM predictions in the total or the differential cross sections, may suggest a new physics signals.
Consequently, the study of the precise theoretical predictions for $\wz$ and $ZZ$ production at the LHC are necessary.

Collaborations at the Tevatron and the LHC have reported experimental results for $\wz$ and $ZZ$ productions, respectively. The measurements for the total and differential cross sections through the leptonic decay mode have been analyzed at the Tevatron~\cite{CDF:2011ab,Abazov:2012cj,D0:2013rca}.
Both ATLAS~\cite{Aad:2011xj,Aad:2012twa,Aad:2011cx,ATLAS-CONF-2014-015,ATLAS-CONF-2013-021,ATLAS-CONF-2013-020} and CMS~\cite{Chatrchyan:2012sga,CMS-PAS-SMP-12-016,CMS-PAS-SMP-13-005,CMS-PAS-SMP-13-011} collaborations have also presented leptonic decay results at the LHC. Besides, $\wz$ and $ZZ$ cross sections, where one $Z$ boson decays to b-tagged jets, have been measured by CMS collaboration~\cite{Chatrchyan:2014aqa}.
However, the data which are currently available from the LHC have large experimental uncertainties~\cite{Wang:2014uea}, so it is worth to make an accurate theoretical predictions beyond QCD NLO to figure out whether the discrepancies come from the theoretical errors or the new physics.
The QCD NLO corrections for the production of $W^{\pm}Z$ and $ZZ$ have been studied in Refs.~\cite{PhysRevD.43.3626,Mele:1990bq,PhysRevD.50.1931,PhysRevD.60.114037,PhysRevD.60.113006,Dixon19983,Frixione:1992pj}.
The transverse momentum resummation for gauge boson pair productions have been calculated in Ref.~\cite{Wang:2013qua}.
In Ref.~\cite{Campanario:2012fk}, $\wz$ production is calculated beyond NLO for high $q_T$ region.
However, Next-to-Next-to-Leading-Order (NNLO) are still in progress.
Very recently, the NNLO corrections to the $ZZ$ production are presented for the LHC~\cite{Cascioli:2014yka,Gehrmann:2013cxs,Henn:2014lfa,Gehrmann:2014bfa,Caola:2014lpa}.
However, in order to compare the total cross section and some kinetic distribution, such as invariant mass distributions, we still need to consider the effects of soft gluon emission near the threshold region. The soft gluon resummation and the approximate NNLO cross sections for $W^+W^-$ pair production were calculated in Ref.~\cite{PhysRevD.88.054028}, and we have repeated their resummation results as a cross check.

In this paper, we calculate the threshold resummation for $\wz$ and $ZZ$ pair production at the Next-to-Next-Leading Logarithmic (NNLL) accuracy based on Soft-Collinear Effective Theory (SCET)~\cite{Beneke:2002ph,Bauer:2000yr,Bauer:2001yt}.
The paper is organized as follows. In Sec.~\ref{s2}, we describe the formalism for threshold resummation in SCET briefly. In Sec.~\ref{s3}, we present the numerical results and some discussion. Then Sec.~\ref{s4} is a brief conclusion.
\section{Factorization and Resummation}\label{s2}
In this section, we briefly review the threshold resummation in SCET formalism in this paper, following the Ref~\cite{Becher:2007ty}. We consider the process
\be
N_1\left(p_1\right)+N_2\left(p_2\right)
   \to V(p_3)+Z(p_4)+X(p_x),
\ee
where $V(=W,Z)$ is a $W$ or $Z$ boson, and $X$ denotes any hadronic final states. In the Born level, the gauge boson pair is mainly produced through $q\bar{q}'$ process:
\be
q(p_1) + \bar{q}'(p_2) \rightarrow V(p_3) + Z(p_4),
\ee
where $p_{i} = z_{i} P_i,~ i=1,2$. We define the kinematic variables as follows
\be
S = (P_1 + P_2)^2,\quad  \hat{s} = (p_1 + p_2)^2,\quad \tau = \mv^2/S,\quad z=\mv^2/\hat{s}.
\ee
where $\mv$ is the invariant mass of the gauge boson pair. During the derivation
of factorization expression, the scale hierarchy is assumed in the threshold region:
\be
\hat{s},\mv^2 \gg \hat{s}(1-z)^2 \gg \Lambda _{\text{QCD}}^2,
\ee
where $\hat{s},\mv^2$ are refered to as hard scales and $\hat{s}(1-z)^2$ is the soft scale. $\lambda=(1-z) \ll 1$ is the  expansion parameter. In the threshold limit, i.e. $\lambda\rightarrow 0$, the cross section can be factorized as
\be
\frac{d\sigma}{d\mv^2} &=& \frac{\sigma_0}{S}\sum _{q,\bar{q}'} \int\frac{dx_1}{x_1
  }\frac{dx_2}{x_2} f_{q}\left(x_1,\mu _f\right)  f_{\bar{q}'}\left(x_2,\mu_f\right)
   \mathcal{H}_{VZ}(\mv,\mu_f)
   \mathcal{S}\left(\sqrt{s}(1-z),\mu _f\right),
\ee
where $\sigma_0$ is the tree level cross section and $\mathcal{S}\left(-\sqrt{s}(z-1),\mu _f\right)$ is the soft function, which given by the vacuum expectation values of soft Wilson loops, while $\mathcal{H}_{VZ}\left(\mv,\mu\right)$ is the hard function, and can be expanded in  powers of $\alpha_s$:
\be
 \mathcal{H}_{VZ}= \mathcal{H}_{VZ}^{(0)} + \frac{\alpha_s}{4\pi}\mathcal{H}_{VZ}^{(1)} + \cdots.
\ee
Here $\mathcal{H}_{VZ}^{(n)}$ can be extracted by matching the perturbative QCD results onto the relevant SCET operator,
and the corresponding complete expression can be found in Ref.~\cite{Wang:2013qua}.

The renormalization-group  equation for the hard function can be written as
\be\label{s2_eq_rg_h}
\frac{d}{d\ln\mu} \mathcal{H}_{VZ}\left(\mv,\mu \right)= 2\left[ \Gamma_{\rm cusp}^F(\alpha_s)\ln\frac{-\mv^2}{\mu^2} + 2\gamma^{q}(\alpha_s)\right]\mathcal{H}_{VZ}\left(\mv,\mu \right),
\ee
There also exist large $\pi^2$ terms in the hard function arising from the negative arguments in the squared logarithmic terms,
which can be resummed to all order if we choose the hard scale as $\mu_h^2\sim-\mv^2$ and then evolve $\mu_h^2$ from the  time-like region to the space-like region. Meantime we need the strong coupling $\alpha_s(\mu^2)$ evaluated at time-like region, and the strong coupling at time-like region $\alpha_s(-\mu^2)$ is related to the running couplings at the space-like region $\alpha_s(\mu^2)$  at NLO by the equation~\cite{Ahrens:2008qu}
\be
\frac{\alpha_s(\mu^2)}{\alpha_s(-\mu^2)}=1-i a(\mu^2)+\frac{\alpha_s(\mu^2)}{4\pi}\left[\frac{\beta_1}{\beta_0}\ln[1-i a(\mu^2)]\right]+\mathcal{O}(\alpha_s^2),
\ee
where $a(\mu^2)=\beta_0\alpha_s(\mu^2)/4$.

The soft function  $\mathcal{S}(s(1-z)^2,\mu)$ can be defined as~\cite{Becher:2007ty}
\be
\mathcal{S}(s(1-z)^2,\mu) = \sqrt{s}W(s(1-z)^2,\mu),
\ee
where the $W(s(1-z)^2,\mu)$ function obeys the form:
\be\label{s2_eq_rg_s}
\omega W(\omega^2,\mu_f) &=&
\exp\left(-4 S\left(\mu _s,\mu _f\right)+2 a_{\gamma ^W}\left(\mu _s,\mu _f\right)\right)
\tilde{s}\left(\partial_\eta,\mu_s\right)\left(\frac{\omega^2}{\mu_s^2}\right)^\eta \frac{e^{-2 \gamma  \eta }}{\Gamma (2 \eta )},
\ee
where $\mu_s$ is the soft scale, $\eta~=~2a_\Gamma(\mu_s,\mu_f)$, and the Sudakov exponent $S$ and the exponents $a_\Gamma$ are defined as
\begin{eqnarray}
 S(\nu,\mu)&=&-\int_{\as(\nu)}^{\as(\mu)}d\alpha \frac{\Gamma_{\rm cusp}^F(\alpha)}{\beta(\alpha)}
 \int_{\as(\nu)}^{\alpha}\frac{d\alpha'}{\beta(\alpha')}, \\
 a_\Gamma(\nu,\mu)&=&-\int_{\as(\nu)}^{\as(\mu)}d\alpha \frac{\Gamma_{\rm cusp}^F(\alpha)}{\beta(\alpha)}.
\end{eqnarray}
The soft Wilson loop under the Laplace transformation is $\tilde{s}(L_s,\mu_s)$. Up to NLO, it can be expressed as
\be
\tilde{s}(L_s,\mu_s) = 1+ \frac{C_F\alpha_s(\mu_s)}{4\pi}\left(2L_s^2+\frac{\pi^2}{3}\right).
\ee

After combining the soft and hard function, the differential cross section can be factorized as
\be\label{s2_eq_nnll}
\frac{d\sigma}{d\mv^2} = \frac{\sigma_0}{S} \int_{\tau }^1 \frac{dz}{z}
\mathcal{L}\left(\frac{\tau }{z},\mu _f\right) \mathcal{H}_{VZ}\left(\mv,\mu _h\right) C\left(\mv,\mu
   _h,\mu _s,\mu _f\right),
\ee
where
\be
\mathcal{L}(y,\mu_f)=\sum_{q,\bar{q}'}\int_y^1 \frac{dx}{x}\left[f_{q}(x,\mu_f)f_{\bar{q}'}(y/x,\mu_f))\right],
\ee
and
$C\left(\mv,\mu _h,\mu _s,\mu _f\right)$ can be written as
\be
C\left(\mv,\mu _h,\mu _s,\mu _f\right)&=&\exp \left[4 S\left(\mu _h,\mu _s\right)-2 a_{\gamma ^V}\left(\mu _h,\mu _s\right)+4
   a_{\gamma ^{\phi }}\left(\mu _s,\mu _f\right)\right] \left(\frac{\mv^2}{\mu _h^2}\right){}^{-2 a_{\Gamma }\left(\mu _s,\mu _h\right)}\nonumber\\
&&\frac{(1-z)^{2 \eta -1}}{z^{\eta }} \tilde{s}\left[\ln \left(\frac{(1-z)^2 \mv^2}{z \mu _s^2}\right)+\partial_\eta,\mu _s\right] \frac{e^{-2 \gamma  \eta }}{\Gamma (2 \eta )}.
\ee

In addition to the singular terms, we should also use contributions from the non-singular terms, which can be obtained by matching resummed results to the full fixed order cross section. Finally, the Renormalization-Group improved prediction for the gauge boson pair production can be expressed as
\begin{eqnarray}\label{s2_eq_fullcs}
\frac{d\sigma^{\rm NNLL + NLO}}{d\mv^2} = \frac{d\sigma^{\rm NNLL}}{d\mv^2} +\left(\frac{d\sigma^{\rm NLO}}{d\mv^2}-\frac{d\sigma^{\rm NNLL}}{d\mv^2}\right)_{\rm expanded~to~NLO}.
\end{eqnarray}

\section{Numerical Results}\label{s3}
In this section, we present the numerical results for the threshold resummation effects on gauge boson pair production. We choose SM input parameters as following~\cite{Beringer:1900zz}:
\begin{eqnarray}\label{sm_para}
 &&  m_W= 80.4 \textrm{~GeV}, \quad    m_Z = 91.19 \textrm{~GeV}, \quad \alpha(m_Z)=1/132.338.
\end{eqnarray}
Throughout the paper, we use MSTW2008nnlo PDFs and associated running QCD coupling constant for the resummation results.
The factorization scale is set as the invariant mass $\mv$.
The fixed-order QCD NLO corrections are calculated by  MCFM~\cite{PhysRevD.60.113006} with MSTW2008nlo PDFs unless specified otherwise, where we consistently choose factorization and renormalization scales as $\mv$.

\subsection{Scale setting and scale uncertanties}
\begin{figure}[t!]
\begin{minipage}[t]{0.45\linewidth}
\centering
  \includegraphics[width=1.0\linewidth]{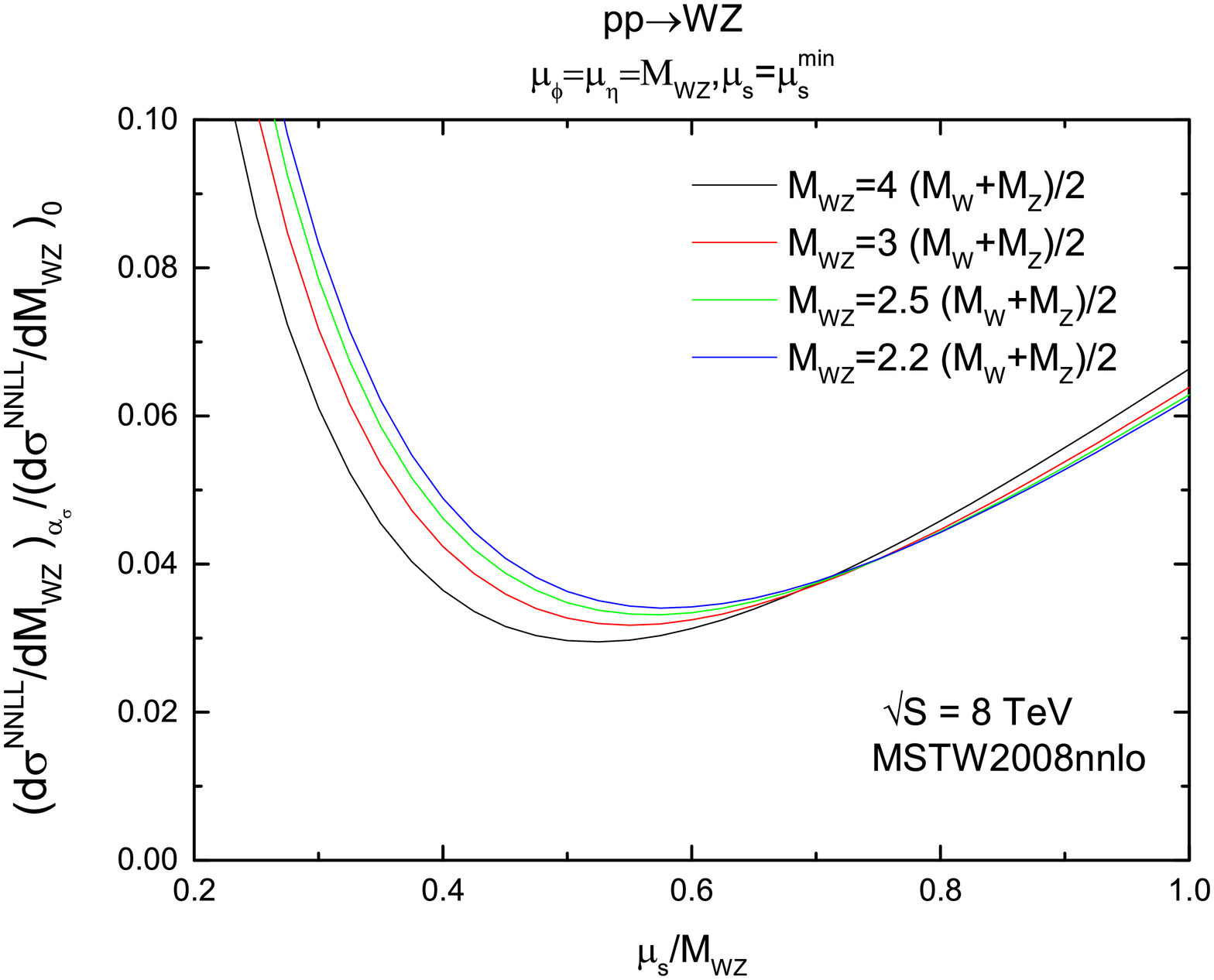}\\
\end{minipage}
\hfill
\begin{minipage}[t]{0.45\linewidth}
\centering
 \includegraphics[width=1.0\linewidth]{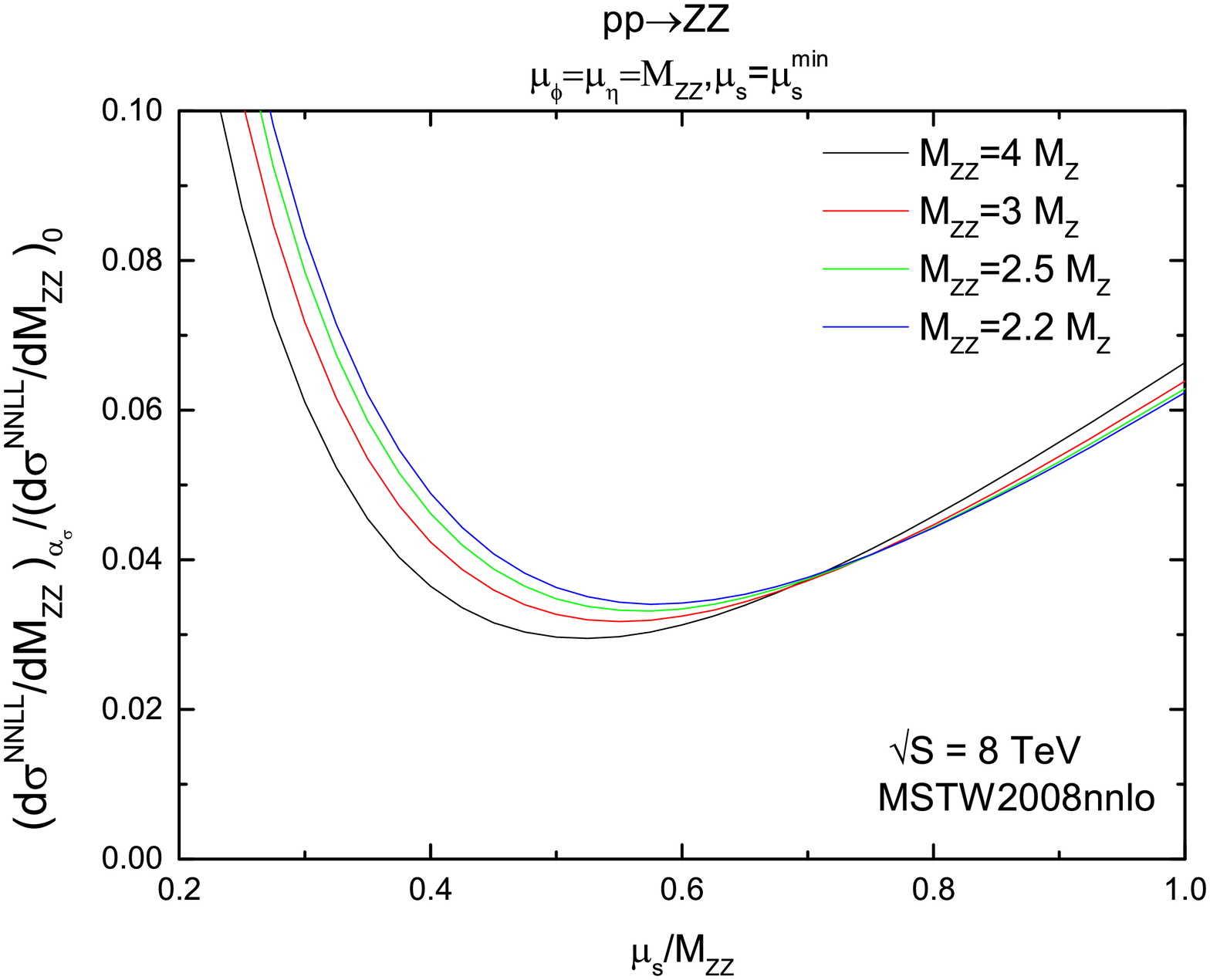}\\
\end{minipage}
\caption{The contribution of the NNLL resummed results arising from the one-loop corrections of the soft functions as a function of $\mu_s/\mv$.}\label{f_soft}
\end{figure}
Before the numerical calculation, two additional scales, the hard scale $\mu_h$ and the soft scale $\mu_s$, also need to be fixed.
Generally, the hard scale should be fixed at $\mu_h\sim\mv$, where the hard Wilson coefficient
have stable perturbative expansions. However, in order to include the $\pi^2$ enhancement effects, we choose $\mu_h^2=-\mv^2$.

In Fig.~\ref{f_soft}, the dependence of the relative corrections of the soft functions at $\mathcal{O}(\alpha_s)$ on the soft scale $\mu_s$ are presented at the LHC with $\eis$.
The requirement on $\mu_s$ is that the soft function should have a well-behaved perturbative stability.
Thus, we determine $\mu_s$ by finding out where the corrections from the soft function is minimized.
Because the soft function $\mathcal{S}(s(1-z)^2,\mu)$ is sensitive to the variable $z$ and to the shape of the PDFs, we should integrate the soft function convoluting PDFs over $z$, and the integration results with different invariant mass are corresponding to the different lines in Fig.~\ref{f_soft}.
As a result, $\mu_s$ is chosen at the minimum point of each line, which can be well  parameterized in the form of
\be
\mu_s = M_{VZ}(1-\tau ) \left(a+b \sqrt{\tau
   }\right)^{-c}.
\ee
The situation with $\fos$ is similar, and not shown here.
Finally, the parameters are chosen as following:
\be\label{s3_eq_s}
\mu_{s,W^{\pm}Z}^{min}= M_{W^{\pm}Z}\frac{1-\tau }{\left(3.004 \sqrt{\tau
   }+1.339\right)^{2.134}},\nno
\mu_{s,ZZ}^{min} = M_{ZZ}\frac{1-\tau }{\left(3.013 \sqrt{\tau
   }+1.323\right)^{2.356}}.
\ee

\begin{figure}[t!]
\begin{minipage}[t]{0.45\linewidth}
\centering
  \includegraphics[width=1.0\linewidth]{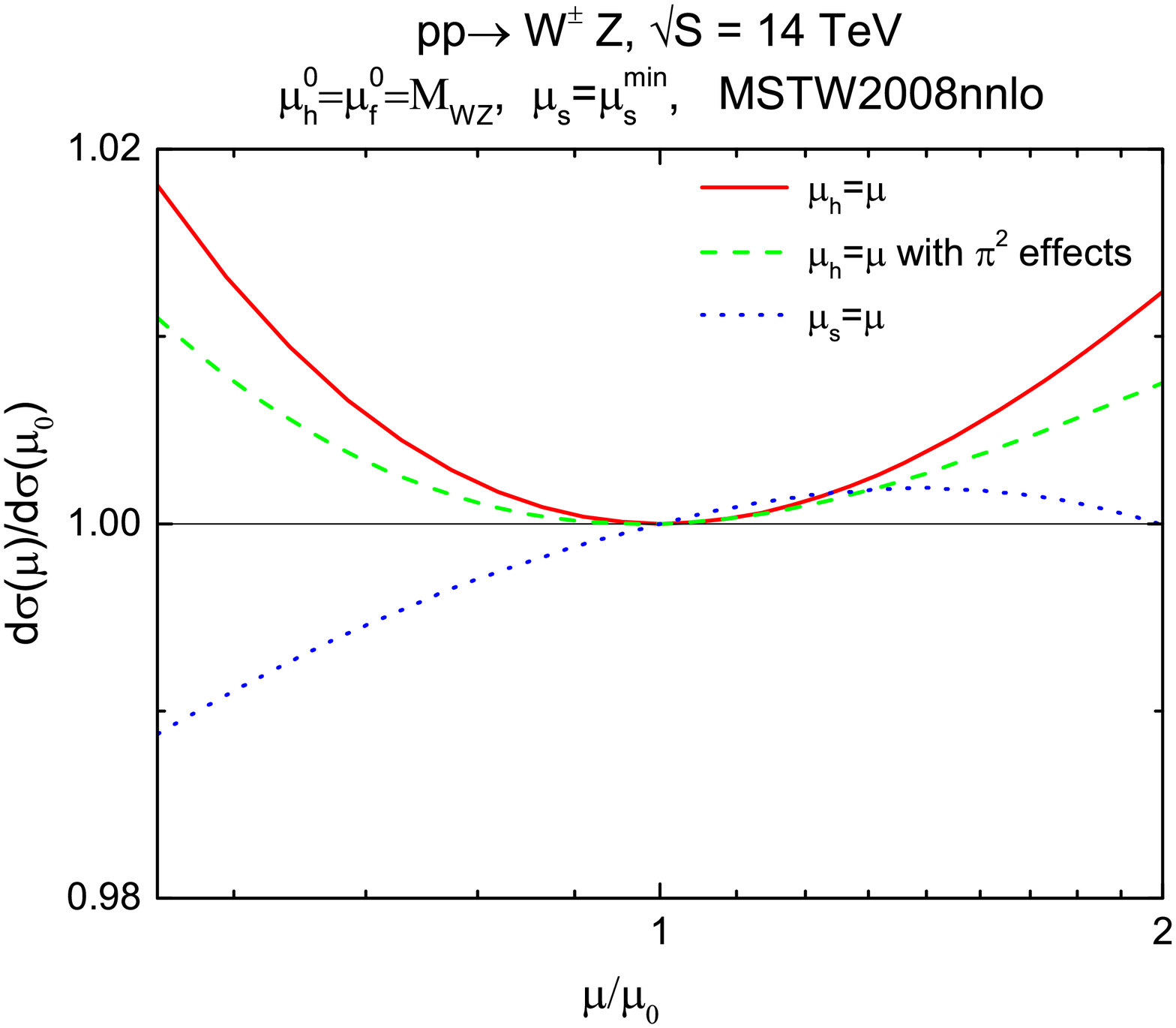}\\
\end{minipage}
\hfill
\begin{minipage}[t]{0.45\linewidth}
\centering
 \includegraphics[width=1.0\linewidth]{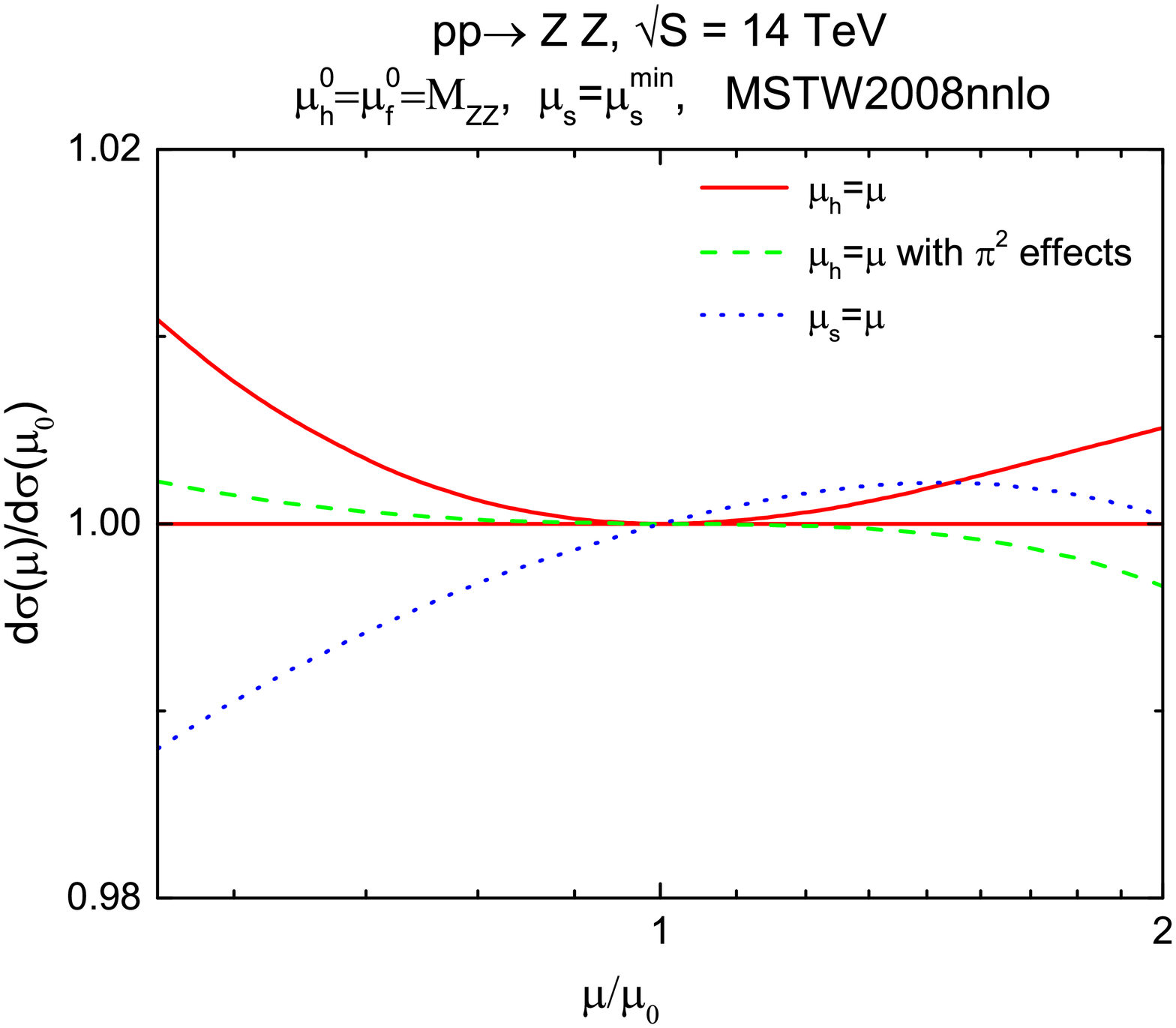}\\
\end{minipage}
\caption{Scale dependence of the resummed cross section on the hard scale, soft scale at NNLL level when $\fos$.}\label{f_sh_scale}
\end{figure}
In Fig.~\ref{f_sh_scale}, we show the scale dependence of the resummed cross sections on the hard scale and the soft scale with $\fos$. They turns out that the scale dependences are very tiny, less than 2\%. In the plots, we also present the results after including $\pi^2$ effects, which decrease the dependence of the hard scale by about $50\%$ comparing with the value without $\pi^2$ effects.

\begin{figure}[t!]
\begin{minipage}[t]{0.45\linewidth}
\centering
 \includegraphics[width=1.0\linewidth]{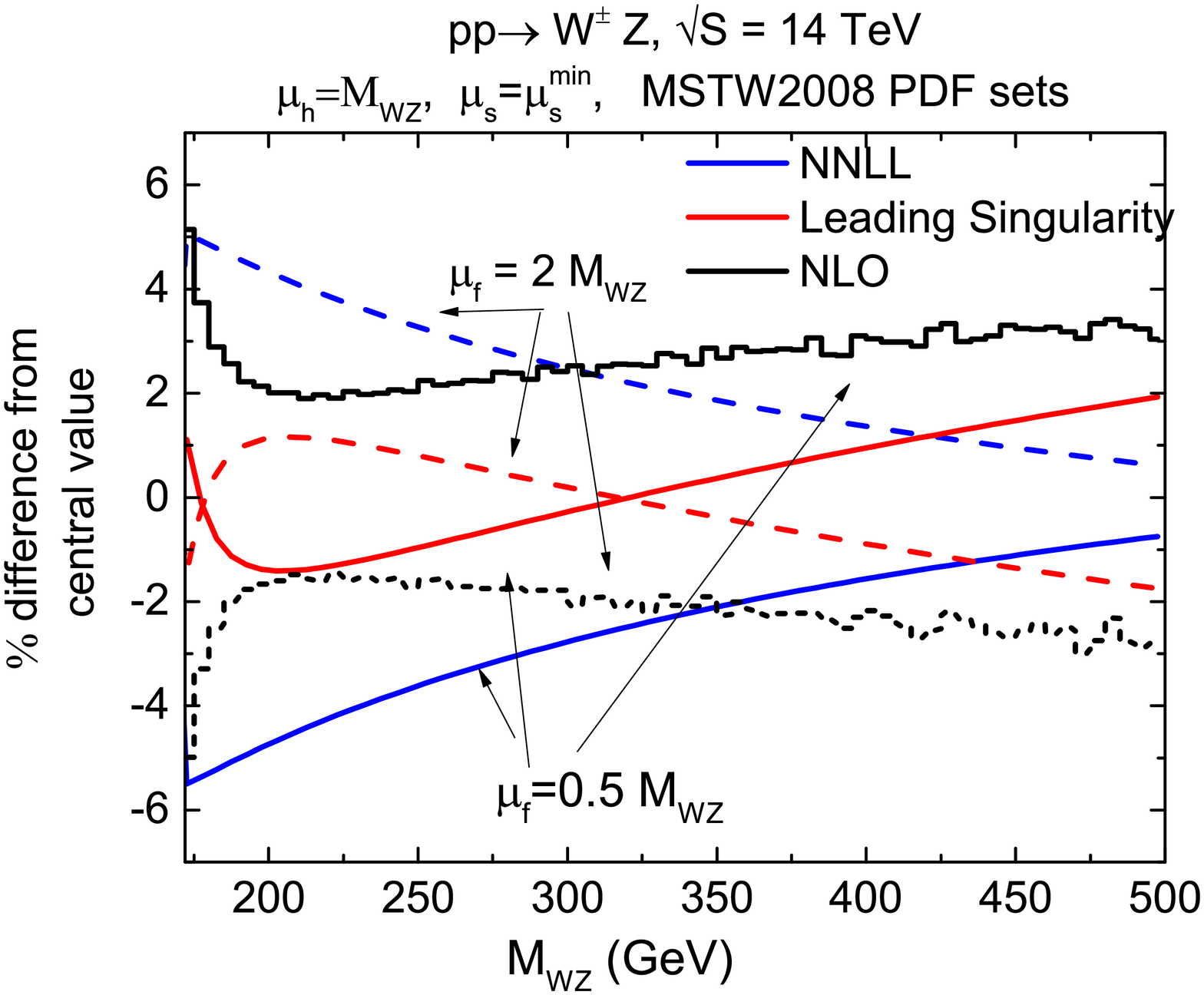}\\
\end{minipage}
\hfill
\begin{minipage}[t]{0.45\linewidth}
\centering
 \includegraphics[width=1.0\linewidth]{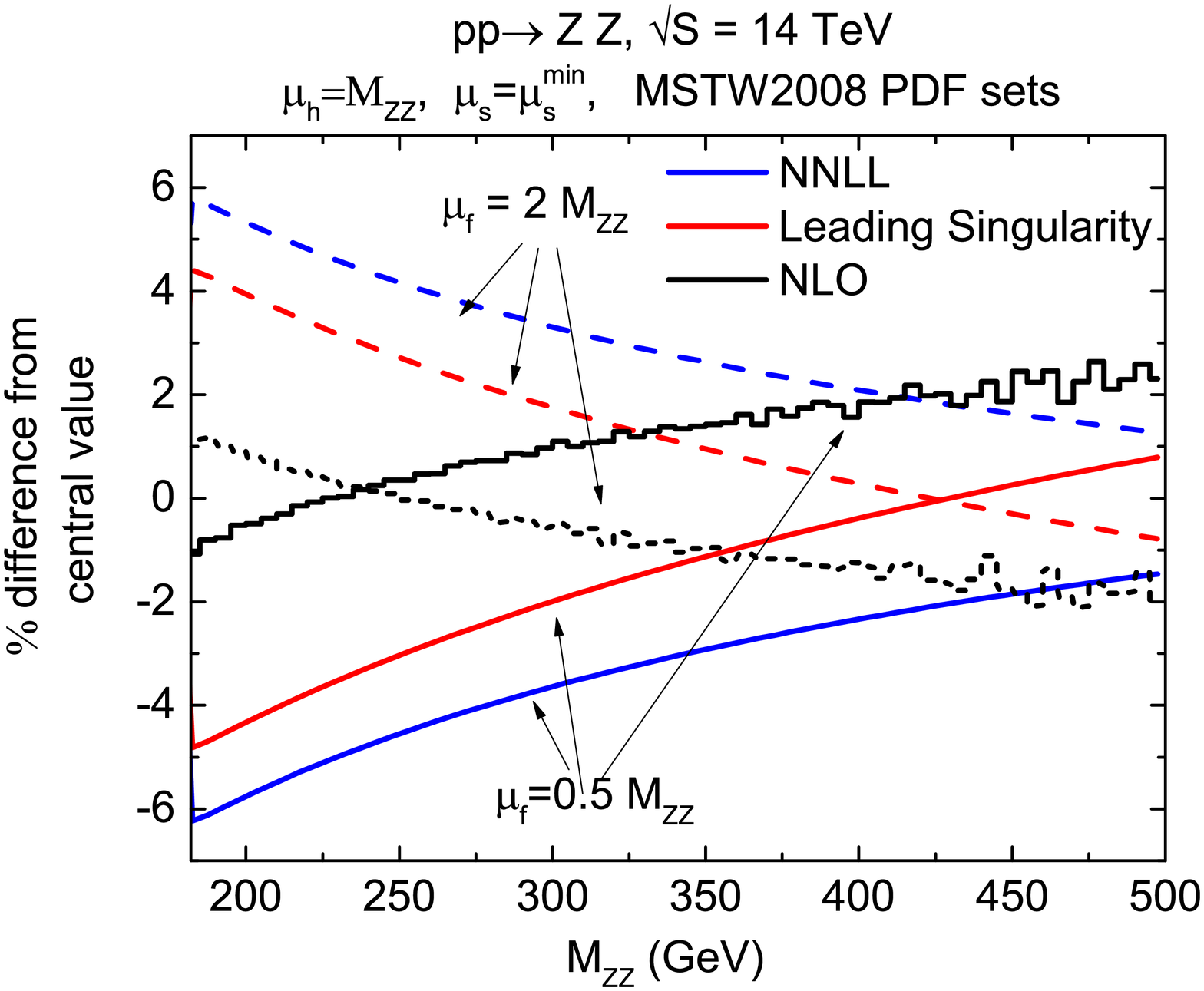}\\
\end{minipage}
\caption{Factorization scale dependence of the NNLL resummed part, the leading singularity part and the fixed-order part  at $\fos$.}\label{f_fscale1}
\end{figure}

\begin{figure}[t!]
\begin{minipage}[t]{0.45\linewidth}
\centering
  \includegraphics[width=1.0\linewidth]{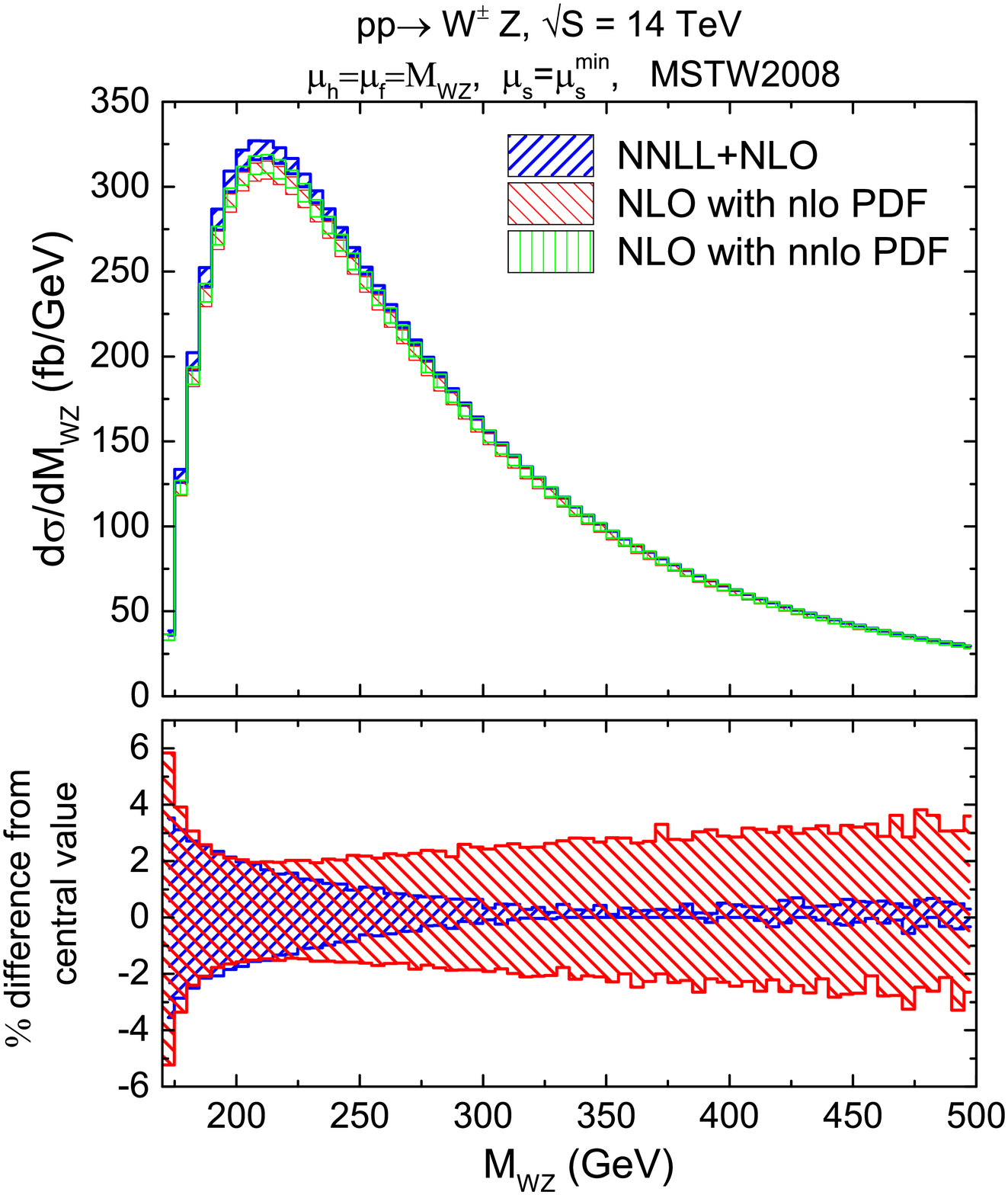}\\
\end{minipage}
\hfill
\begin{minipage}[t]{0.45\linewidth}
\centering
  \includegraphics[width=1.0\linewidth]{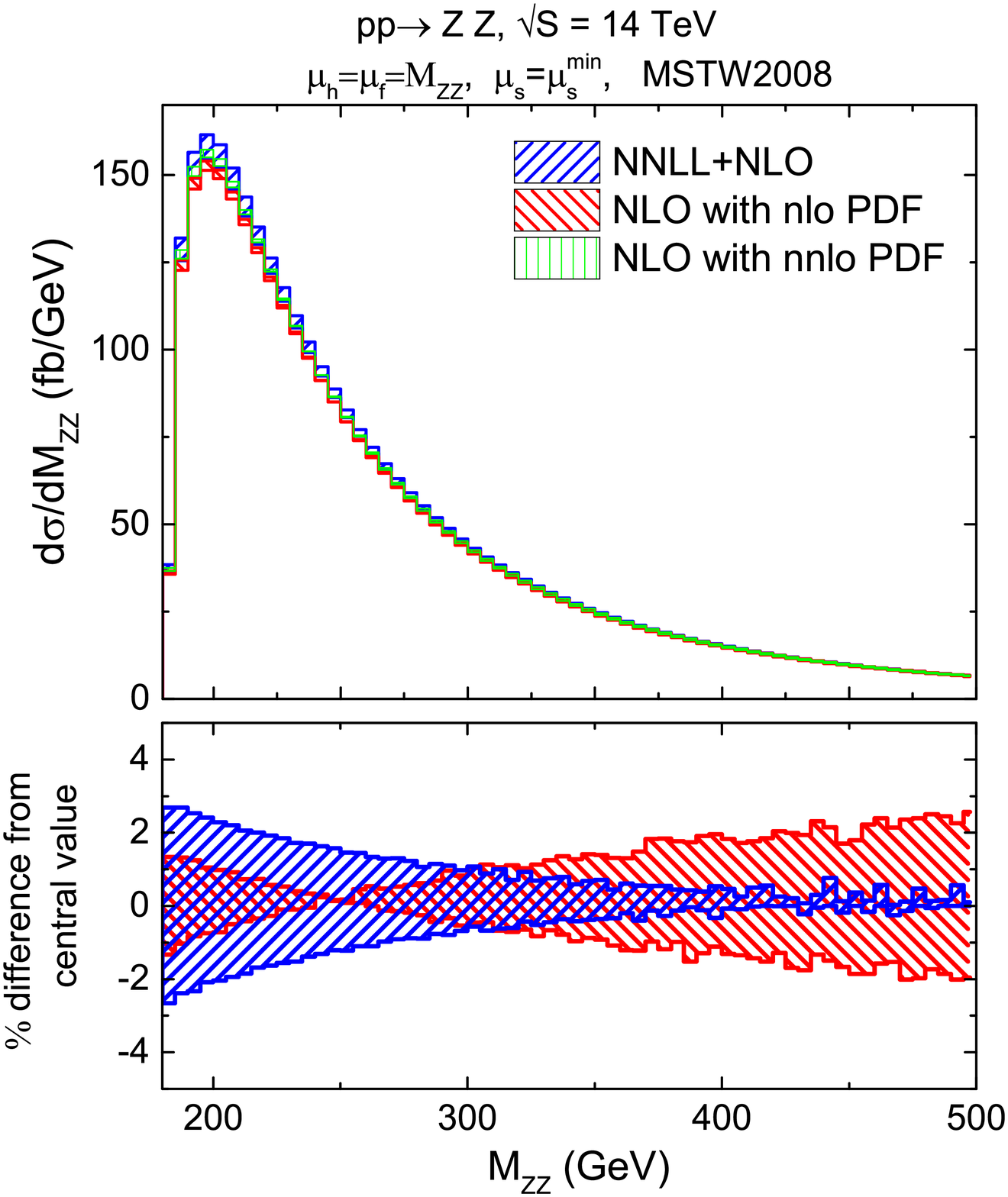}\\
\end{minipage}
\caption{Comparison of NNLL+NLO resummation results with NLO predictions, and their factorization scale uncertainties from respective central values at $\fos$. The Blue lines are the NNLL+NLO resummation results, the red lines are the NLO prediction with nlo PDF, while the green lines are those with nnlo PDF.}\label{f_fscale2}
\end{figure}

Fig.~\ref{f_fscale1} and Fig.~\ref{f_fscale2} show the factorization scale dependences on the invariant mass for $\wz$ and $ZZ$ production.
In Fig.~\ref{f_fscale1}, we compare the factorization scale dependence of the resummed part, the leading singularity part and the fixed-order part in Eq~(\ref{s2_eq_fullcs}), where $\mu_f$ are changed from $\mv/2$ to $2\mv$.
The dependence are defined as the ratio of their respective central value.
We find that the scale dependence of the resummed part and the leading singularity part have the same tendency, while that of the NLO part is opposite to the resummed part, so that they cancel each other according to Eq.~(\ref{s2_eq_fullcs}).

The full factorization scale dependences on the invariant mass are described in Fig.~\ref{f_fscale2} (up).
The blue bands correspond to the NNLL+NLO resummation results,
and the red bands are the NLO results with nlo PDFs.
The NLO predictions with nnlo PDFs are also shown as the green bands.
Comparing with the green and the blue bands, we can find that the main corrections come from the resummation effects, but not the PDFs.
Fig.~\ref{f_fscale2} (down) are the deviation from the results of the NNLL+NLO resummation and NLO predictions with nlo PDFs at the central scales, respectively.
The NLO results with nnlo PDFs are not shown here, because they are almost the same as the results with nlo PDFs.
We can see that in the large invariant mass region, the $\mu_f$ dependences of the NNLL+NLO resummation results are much smaller than those of the NLO results.
Besides, the $\mu_f$ dependences of the $\wz$ production are smaller than that of the $ZZ$ production.
Because there are larger scale dependences in the $\wz$ fixed-order term,
which can cancel out more scale dependences in the resummed part as mentioned above, than the case of the $ZZ$ production.

\begin{figure}[t!]
\begin{minipage}[t]{0.45\linewidth}
\centering
 \includegraphics[width=1.0\linewidth]{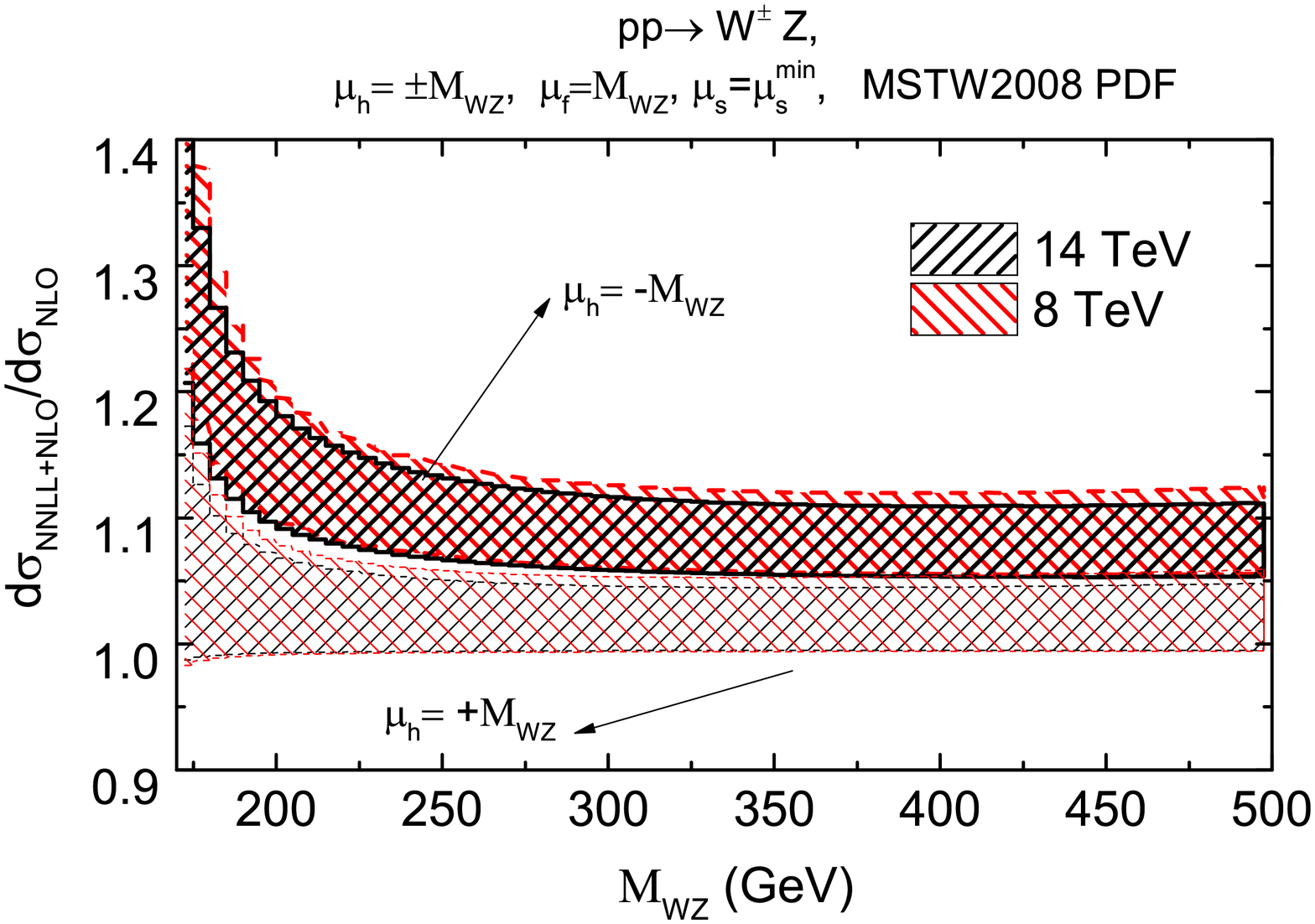}\\
\end{minipage}
\hfill
\begin{minipage}[t]{0.45\linewidth}
\centering
 \includegraphics[width=1.0\linewidth]{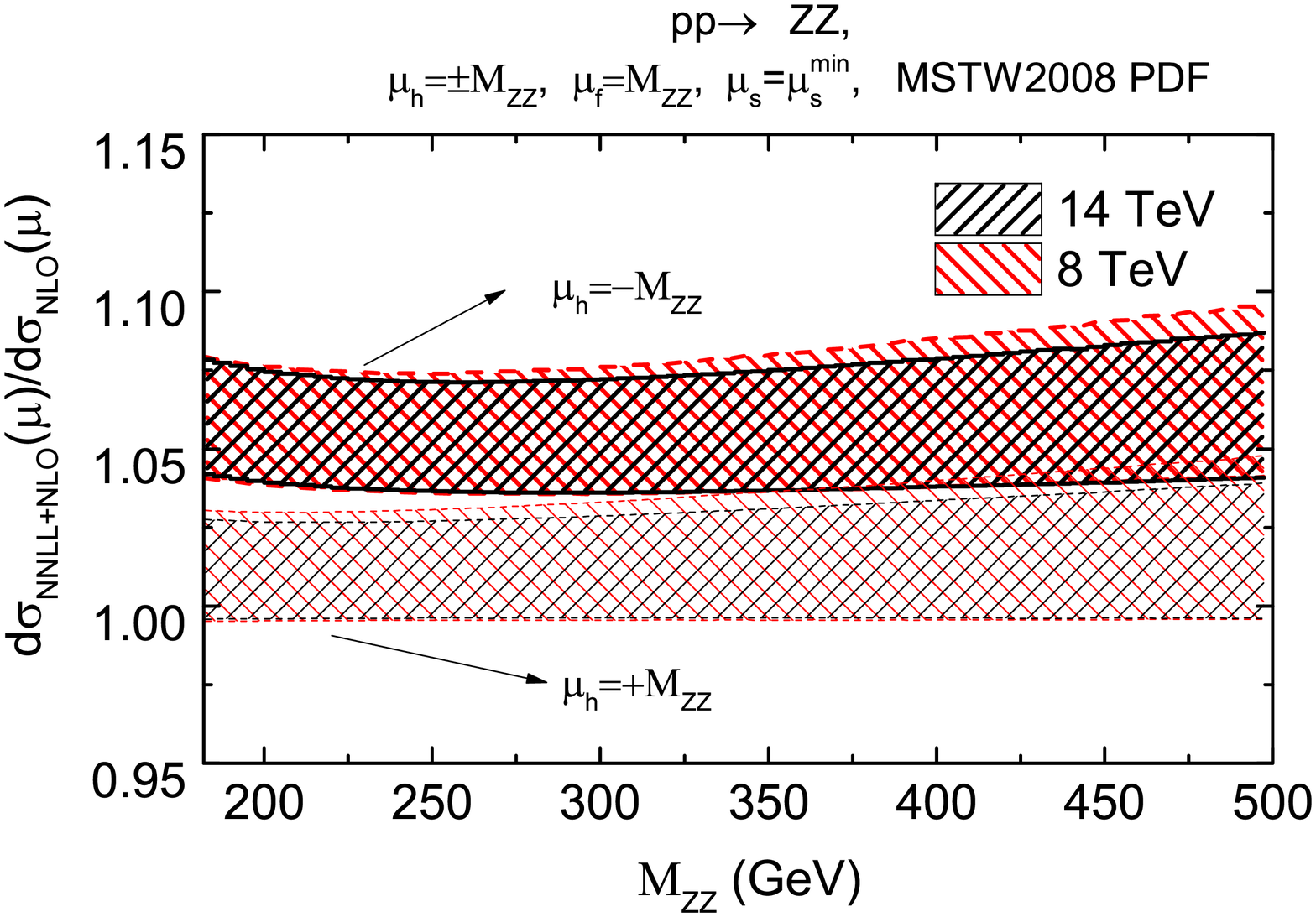}\\
\end{minipage}
\caption{The ratio of the NNLL+NLO results, with and without the $\pi^2$ enhancements effects, to the NLO predictions.}\label{f_kfactor}
\end{figure}
The ratio of the NNLL+NLO prediction, with (without) $\pi^2$ enhancements effects, to the NLO results are shown in the Fig.~\ref{f_kfactor}. For dash bands, only the regular soft gluon resummation effects are included, while for solid bands the $\pi^2$ enhancements effects are also included. For the both cases, in the large invariant mass region, the corrections are stable, and about $9\%$ ($2\%$) for $\wz$ production and $6\%$ ($1\%$) for $ZZ$ production with (without) the $\pi^2$ enhancements effects, respectively.
The large corrections in the low invariant mass region are mainly due to the $\pi^2$ enhancement effects, especially for $\wz$ production. The contributions of the $\pi^2$ enhancement effects for the gauge boson pair are similar to those found for Drell-Yan process, which are proportional to the $\alpha_s(\mu_h^2)$ at LO~\cite{Ahrens:2008qu,Shao:2013bz}.  Thus, the corrections become large at small invariant mass. In the following, we will always include the effects of the $\pi^2$ resummation.
\subsection{Invariant mass distribution and total cross sections}

\begin{figure}[t!]
\begin{minipage}[t]{0.45\linewidth}
\centering
 \includegraphics[width=1.0\linewidth]{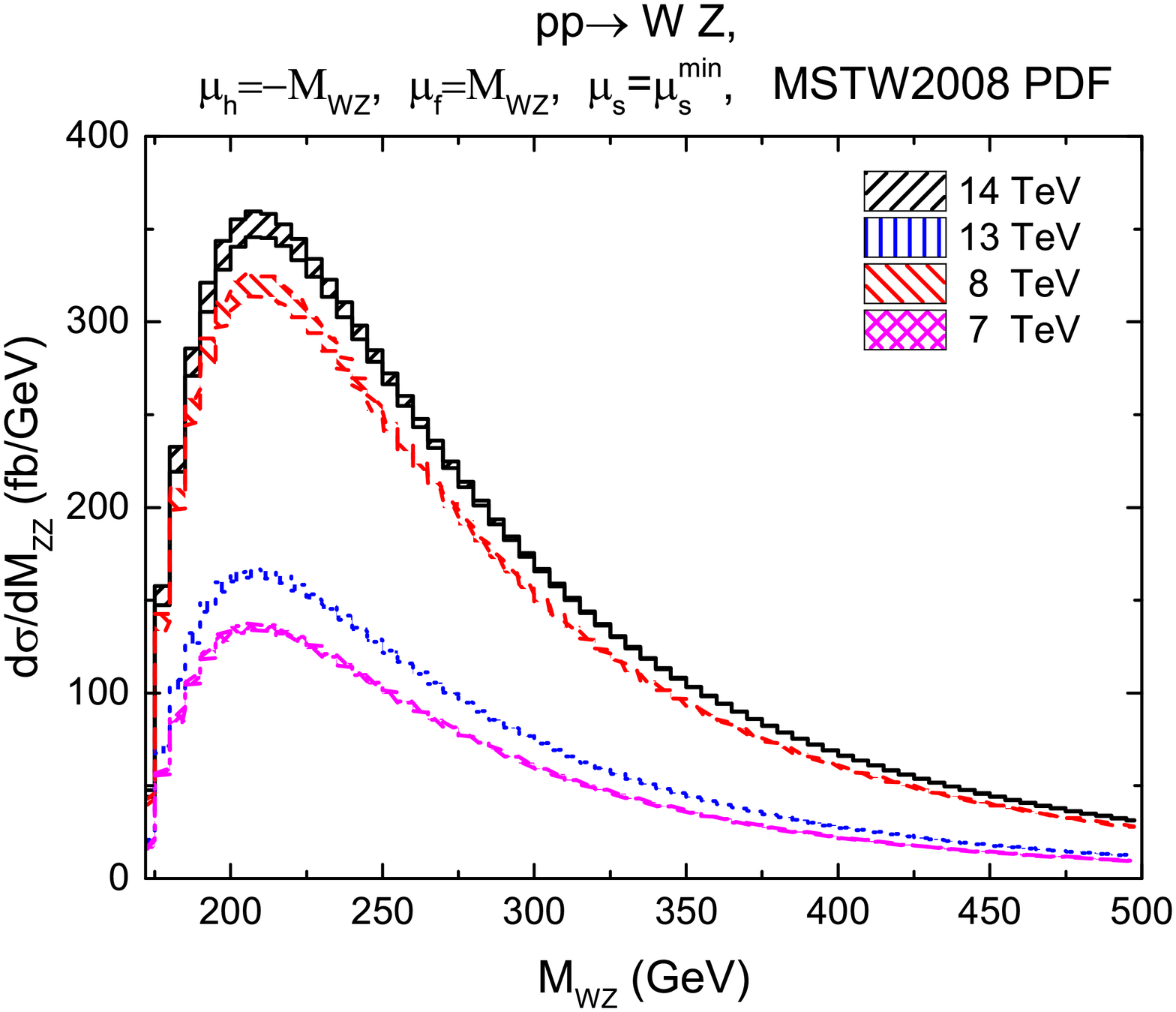}\\
\end{minipage}
\hfill
\begin{minipage}[t]{0.45\linewidth}
\centering
 \includegraphics[width=1.0\linewidth]{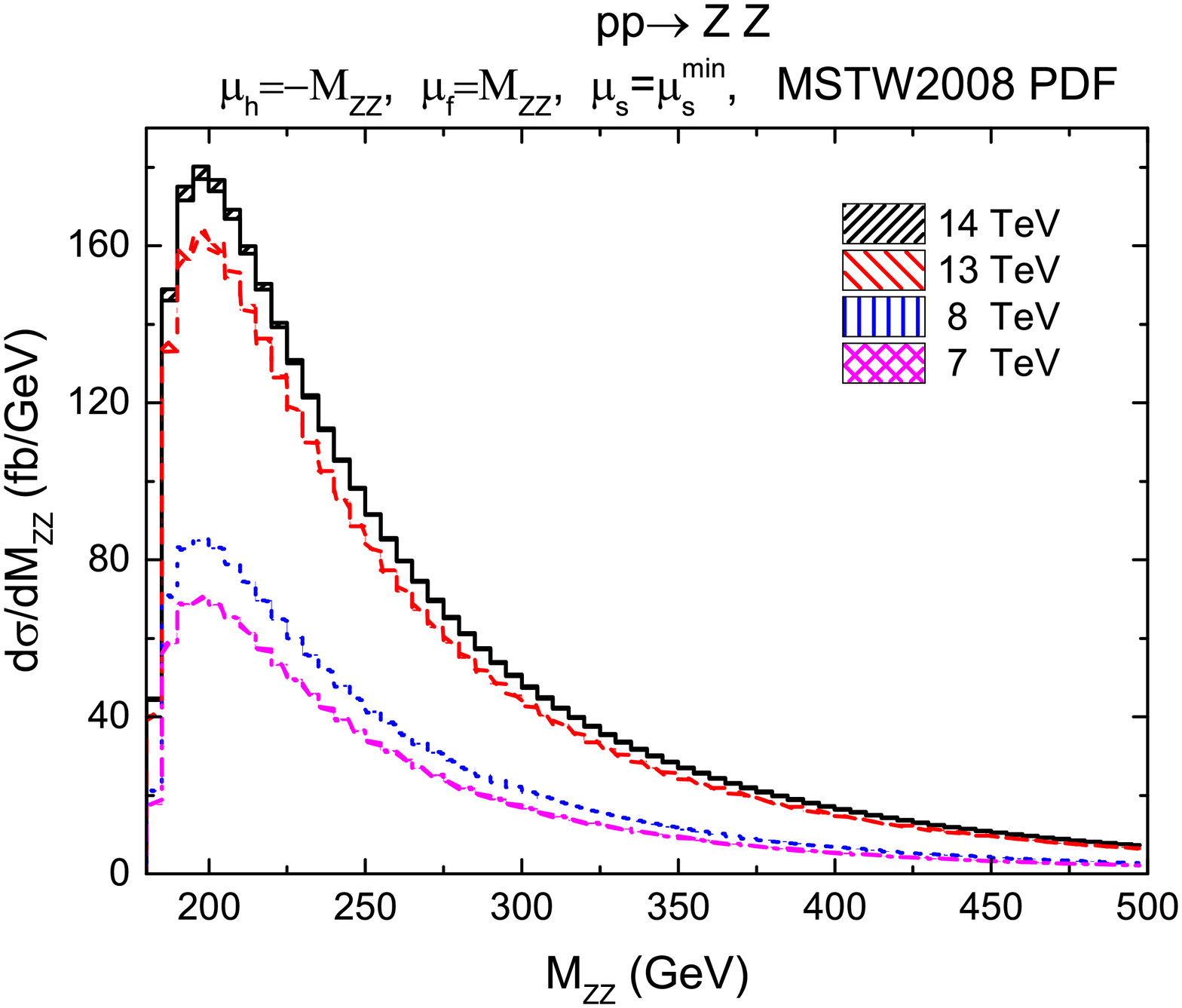}\\
\end{minipage}
\caption{The invariant mass distributions with $\sqrt{S} = 7,~8,~13$ and $14$ TeV, and the scale is $\mu_h^2=-\mv^2$ and $\mu_s=\mu_s^{min}$.}\label{f_energy}
\end{figure}
In Fig.~\ref{f_energy}, the full invariant mass distribution for $\wz$ and $ZZ$ production for various energies are shown. We also include contributions from gluon initial states in $ZZ$ productions. The peak position are at about 210 GeV for $W^{\pm}Z$ production and 200 GeV for $ZZ$ production, respectively. With the increasing of the collider energy, the peak position moves a little to the high invariant mass region.

\begin{figure}[t!]
\begin{minipage}[t]{0.45\linewidth}
\centering
\includegraphics[width=1\linewidth]{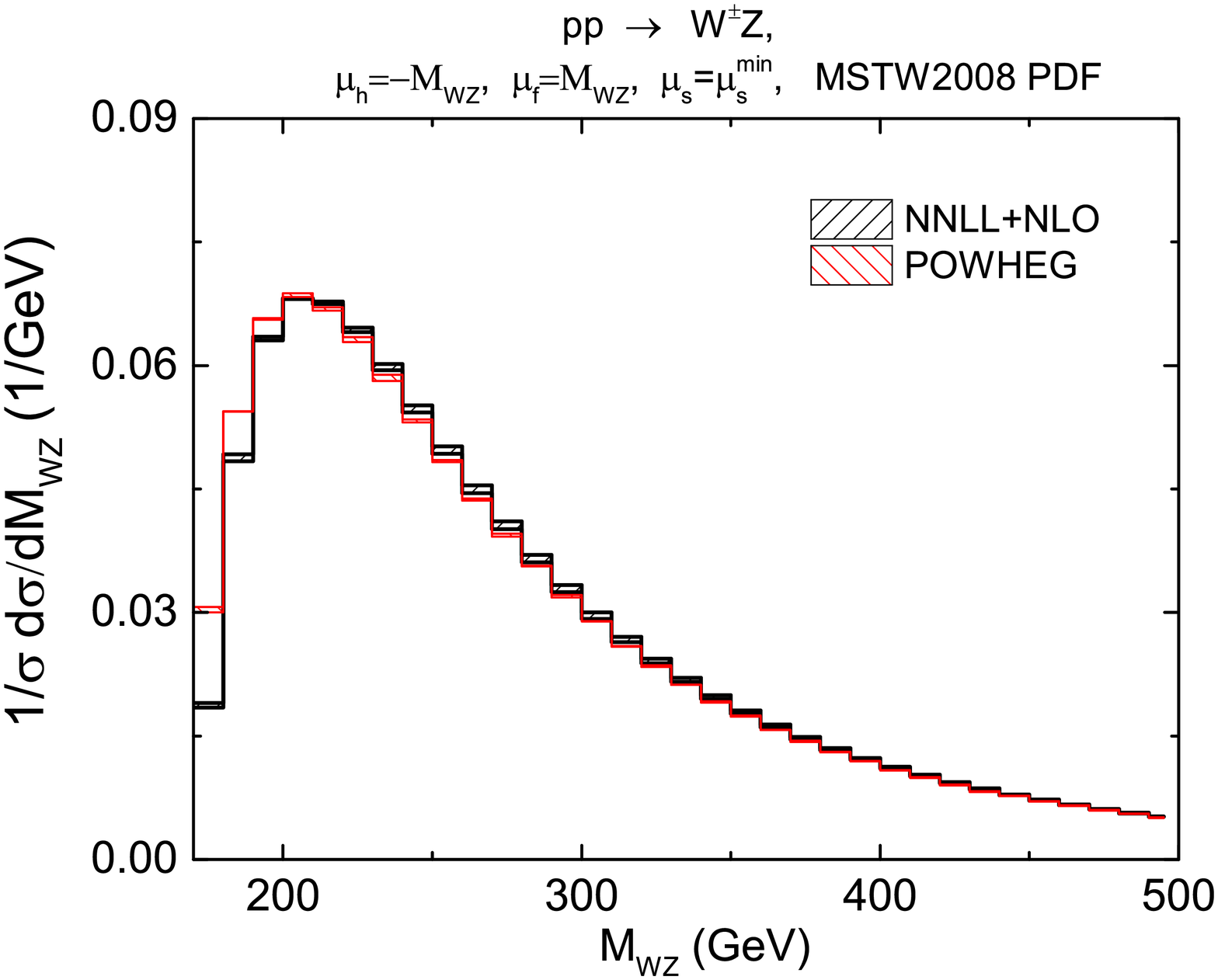}\\
\end{minipage}
\hfill
\begin{minipage}[t]{0.45\linewidth}
\centering
\includegraphics[width=1\linewidth]{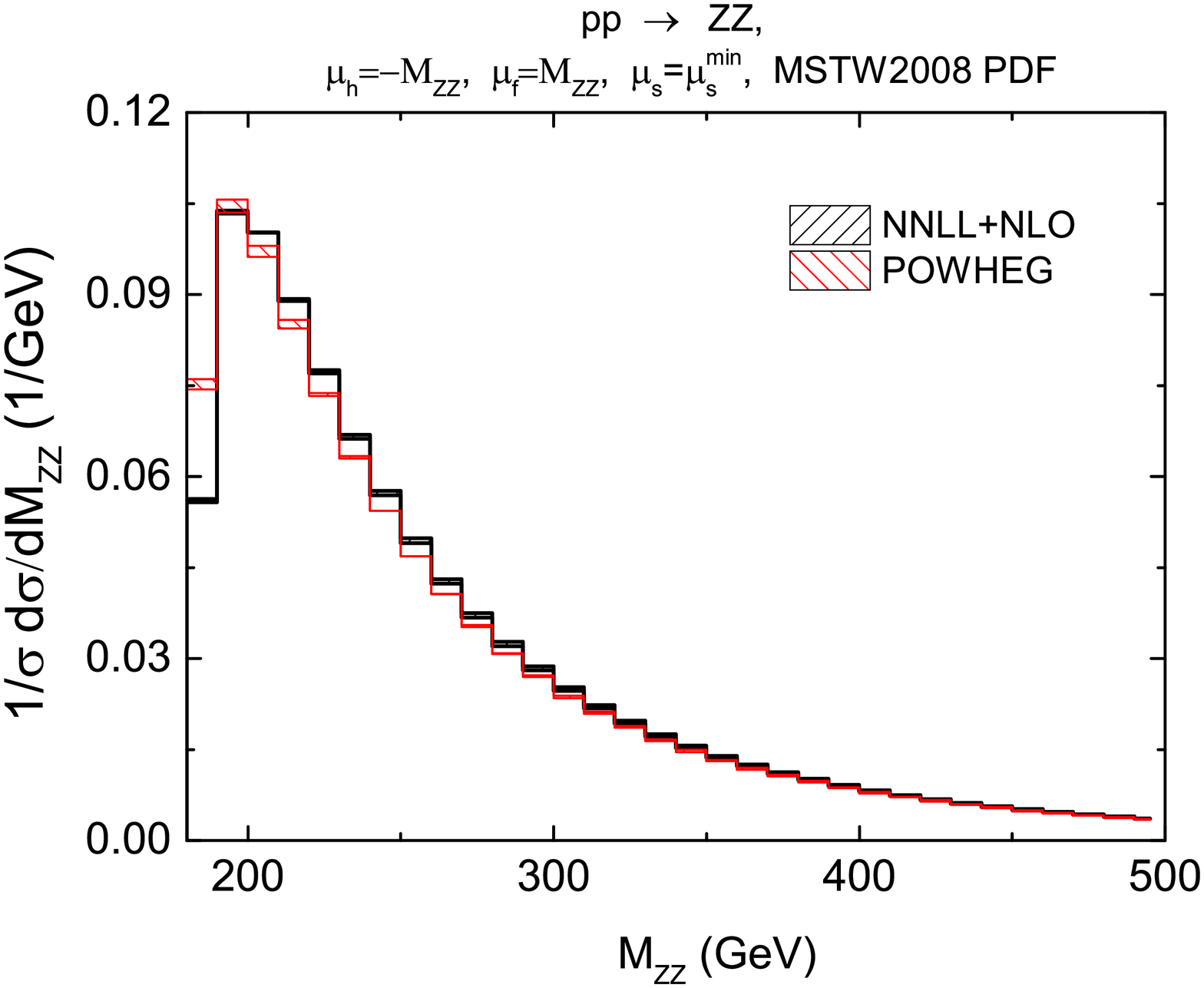}\\
\end{minipage}
  \caption{Comparison of normalized invariant mass distribution for $W^{\pm}Z$ and $ZZ$ productions between POWHEG and resummation predictions at the LHC with $\eis$.}\label{f_exp}
\end{figure}
In Fig.~\ref{f_exp}, we compare the results of the NNLL+NLO resummation and the POWHEG~\cite{Nason:2013ydw} at $\eis$, where the off-shell effects, the singly resonant contribution and interference with identical fermions are ignored in POWHEG. Both results agree with each other in most regions. The slight differences mainly lie in the peak region.

In Table~\ref{t_crox1}, we summarize the total cross sections for $\wz$ production at the LHC.
The first row is the NLO cross sections for $q\bar{q}'$ initial states.
The second row is the NNLL+NLO resummation predictions.
The third row is the resummation results including the $\pi^2$ enhancement effects.

Table~\ref{t_crox2} are the results for $ZZ$ production.
$\sigma^{\rm{gg}}$ are the $gg$ channel contributions, which are calculated with nlo PDFs.
And $\sigma^{\rm{NNLL+NLO}}_{\rm{tot}}$ are the total cross sections of NNLL+NLO resummation, including NLO $gg$ contributions.
Comparing with the $ZZ$ production, the $\pi^2$ enhancement effects are significant for $\wz$ production,
which come from the differences of the scale independent term in the hard function between the two channels~\cite{Mele:1990bq,Frixione:1992pj}.

In Table~\ref{t_crox1} and~\ref{t_crox2}, the uncertainties arise from varying the hard, soft scales and the factorization scale each separately by a factor of two around the default choice.
These uncertainties are added up in quadrature.
The uncertainties of the resummation results increase with the center-of-mass energy.
For $\wz$ production, the uncertainties for $\sigma^{\rm{NNLL+NLO}}_{\pi^2}$ are much better than $\sigma^{\rm{NLO}}$ in any center-of-mass energy. For $ZZ$ production at $\sqrt{S}=13$, and 14 TeV, although the factorization scale dependences for $\sigma^{\rm{NNLL+NLO}}_{\pi^2}$ are less than those for $\sigma^{\rm{NLO}}$, after taking the soft and hard scale variation into account, the uncertainties become a little greater than that of $\sigma^{\rm{NLO}}$.

\begin{table}[h]
\begin{center}
\begin{tabular}{|c|c|c|c|c|}
  \hline
  $\sigma$ (pb) & $\ses$ & $\eis$ & $\ths$ & $\fos$ \\
  \hline
  $\sigma^{\rm{NLO}}$              & $17.28_{-0.52}^{+0.65}$         & $21.37^{+0.76}_{-0.61}$     &$44.16^{+1.20}_{-0.94}$     &
  $49.09^{+1.27}_{-0.10}$      \\

  $\sigma^{\rm{NNLL+NLO}}$         & $17.88^{+0.43}_{-0.22}$       & $22.10^{+0.53}_{-0.27}$ & $45.69^{+1.07}_{-0.58}$  & $50.77^{+1.20}_{-0.65}$  \\

  $\sigma^{\rm{NNLL+NLO}}_{\rm{\pi^2}}$ & $19.40^{+0.30}_{-0.24}$       & $23.96^{+0.37}_{-0.30}$  & $49.35^{+0.83}_{-0.68}$ & $54.81^{+0.94}_{-0.77}$\\
  \hline
\end{tabular}
\end{center}
  \caption{Total cross sections for $pp \rightarrow W^{\pm}Z$ with MSTW2008 PDFs.}\label{t_crox1}
\end{table}

\begin{figure}[t!]
\begin{minipage}[t]{0.45\linewidth}
\centering
 \includegraphics[width=1.0\linewidth]{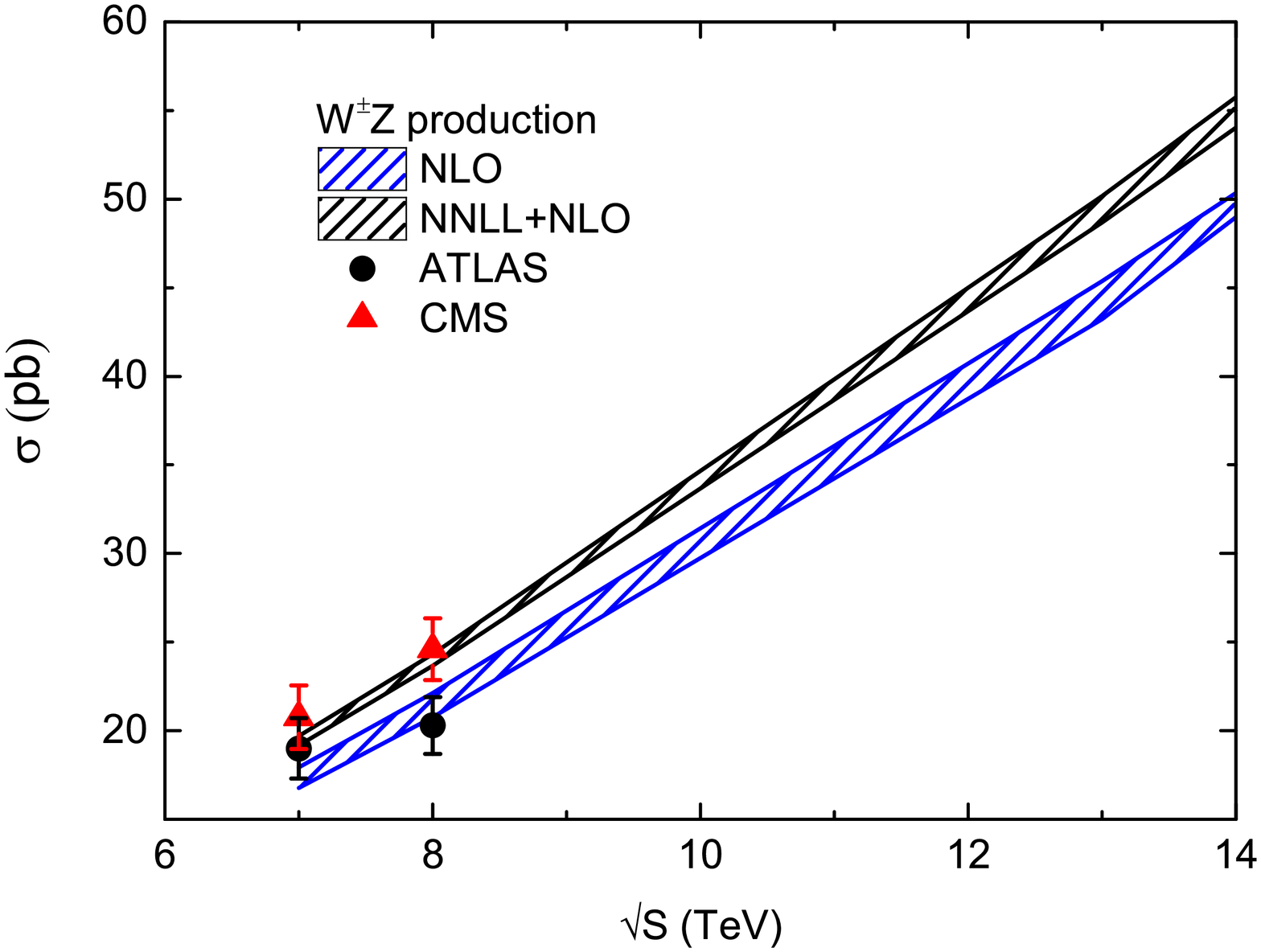}\\
\end{minipage}
\hfill
\begin{minipage}[t]{0.45\linewidth}
\centering
 \includegraphics[width=1.0\linewidth]{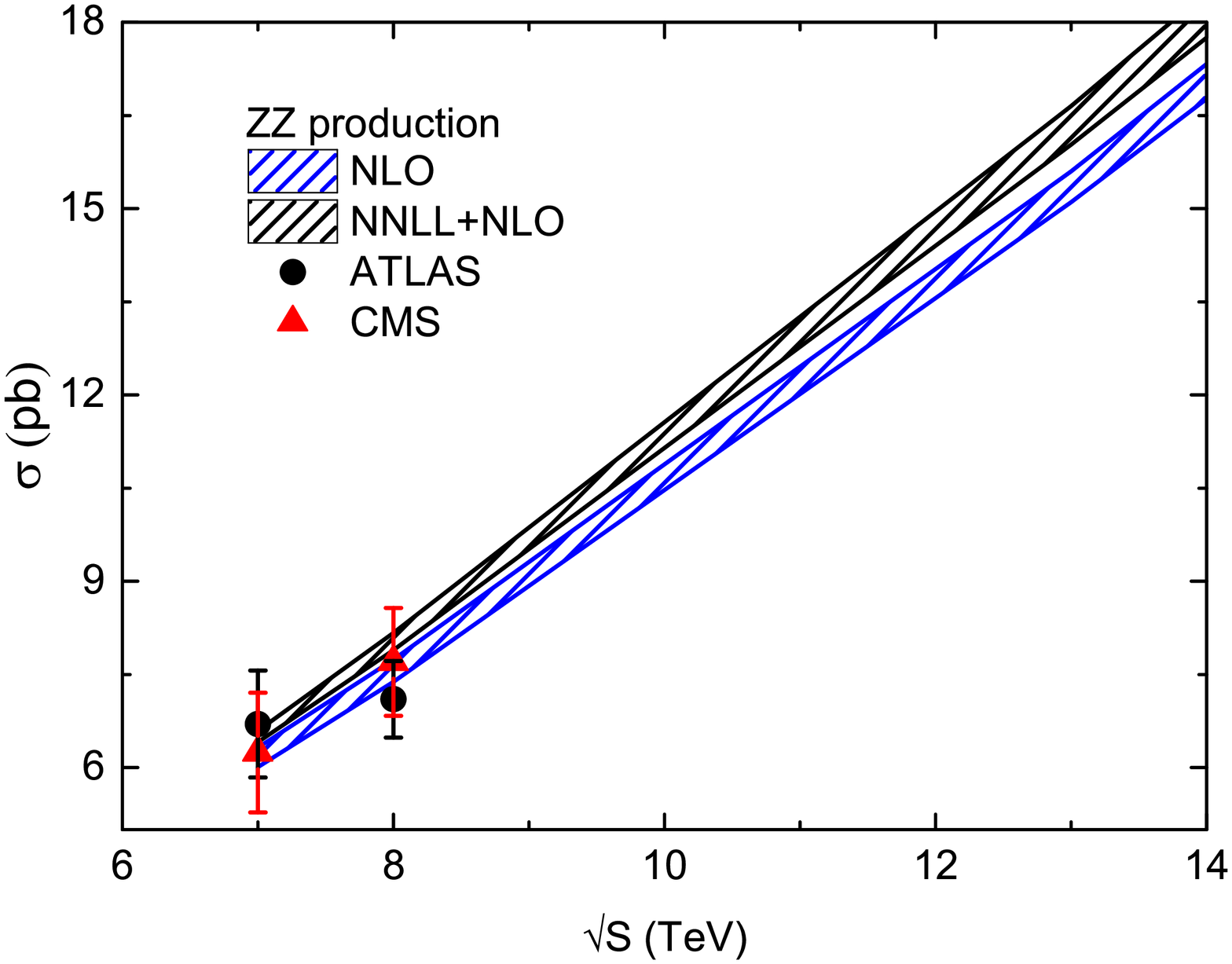}\\
\end{minipage}
\caption{The total cross sectiones with different  center-of-mass energies for gauge boson pair production at the LHC.}\label{f_exp_tot}
\end{figure}

 \begin{table}[h]
\begin{center}
\begin{tabular}{|c|c|c|c|c|}
  \hline
  $\sigma$ (pb) & $\ses$ & $\eis$ & $\ths$ & $\fos$ \\
    \hline
  $\sigma^{\rm{NLO}}$              & $5.86^{+0.10}_{-0.07}$      & $7.16^{+0.10}_{-0.07}$     & $14.26^{+0.08}_{-0.02}$      & $15.77^{+0.07}_{-0.01}$ \\
  $\sigma^{\rm{gg}}$               & $0.28^{+0.08}_{-0.06}$      & $0.38^{+0.1}_{-0.09}$       & $1.06^{+0.24}_{-0.20}$      & $1.22^{+0.27}_{-0.21}$ \\

  $\sigma^{\rm{NNLL+NLO}}$         & $5.98^{+0.08}_{-0.07}$ & $7.33^{+0.10}_{-0.10}$   & $14.66^{+0.27}_{-0.24}$  & $16.21^{+0.31}_{-0.27}$ \\

  $\sigma^{\rm{NNLL+NLO}}_{\rm{\pi^2}}$ & $6.25^{+0.04}_{-0.08}$ & $7.65^{+0.11}_{-0.11}$ & $15.31^{+0.23}_{-0.25}$ & $16.94^{+0.27}_{-0.30}$\\

  $\sigma^{\rm{NNLL+NLO}}_{\rm{tot}}$   & $6.53^{+0.09}_{-0.10}$ & $8.03^{+0.15}_{-0.14}$ & $16.37^{+0.33}_{-0.32}$ & $18.16^{+0.38}_{-0.37}$\\
  \hline
\end{tabular}
\end{center}
  \caption{Total cross sections for $pp \rightarrow ZZ$ with MSTW2008 PDFs.}\label{t_crox2}
\end{table}
In Fig.~\ref{f_exp_tot}, we summarize and compare the total cross section with the LHC experiment data. Obviously, within theoretical and experimental uncertainties, our NNLL+NLO predictions are consistent with the experimental data~\cite{Wang:2014uea}. The largest deviation is less than 2 $\sigma$  as compared to ATLAS data at $\eis$.

\section{Conclusion}\label{s4}
We have calculated the threshold resummation for $W^{\pm}Z$  and  $ZZ$ pair productions at the NNLL + NLO accuracy at the LHC with SCET.
We present the invariant mass distributions and the total cross sections, including $\pi^2$ enhancement effects, which show that the resummation effects increase the NLO total cross section by about 7\% for $ZZ$ production and 12\% for $\wz$ production, respectively.
Our results also agree well with the experimental data reported by ATLAS and CMS collaboration both at $\ses$ and at $\eis$ within theoretical and experimental uncertainties.
\section{ACKNOWLEDGMENTS}
This work is supported in part
by the National Natural Science Foundation of China
under Grants No. 11375013 and No. 11135003.
\bibliography{wz}
\end{document}